\documentclass[journal]{IEEEtran}

\usepackage{ifpdf}
 \ifpdf
 \else
 \fi

\usepackage{cite}

\ifCLASSINFOpdf
   \usepackage[pdftex]{graphicx}
   \usepackage{epstopdf}
   \usepackage{subcaption}
\else
   \usepackage[dvips]{graphicx}
\fi
\usepackage[cmex10]{amsmath}
\usepackage{amssymb}
\usepackage{amsthm}
\usepackage{hyperref}
\begin{document}
%
\title{Block-Level Unitary Query: Incorporating Orthogonal-like Space-time Code with Query Diversity for MIMO Backscatter RFID}
%
%
%

\author{Chen~He,~\IEEEmembership{Member,~IEEE},~
        Z.~Jane~Wang,~\IEEEmembership{Senior Member,~IEEE},~
        and Victor~C.M.~Leung, ~\IEEEmembership{Fellow,~IEEE}
        \thanks{The authors are with department of Electrical and Computer Engineering, the University of British Columbia, Vancouver, Canada. Emails \{chenh, zjanew, vleung\}@ece.ubc.ca.}
}

%
%

\markboth{}%
{Shell \MakeLowercase{\textit{et al.}}: Bare Demo of IEEEtran.cls for Journals}
%



\maketitle


\begin{abstract}
Because of the emerging field of Internet of Things (IoT),  future backscatter RFID is required to be more reliable and data intensive. Motivated by this, orthogonal space-time block code (OSTBC), which is very successful in mobile communications for its low complexity and high performance, has already been investigated for backscatter RFID. On the other hand, a recently proposed scheme called unitary query was shown to be able to considerably improve the reliability of backscatter radio by exploiting query diversity. Therefore incorporating the classical OSTBC (at the tag end) with the recently proposed unitary query (at the query end) seems to be promising.
However, in this paper, we show that simple, direct employment of OSTBC together with unitary
query incurs a linear decoding problem and eventually leads to a severe performance degradation. As a re-design of the recently proposed unitary query and the classical OSTBC specifically for MIMO backscatter RFID,  we present a BUTQ-mOSTBC design pair idea by proposing the block-level unitary query (BUTQ) at the query end and the corresponding modified OSTBC (mOSTBC) at the tag end. The proposed BUTQ-mOSTBC can resolve the linear decoding problem, keep the simplicity and high performance properties of the classical OSTBC, and achieve the query diversity for the $M \times L \times N$ MIMO backscatter RFID channel.
\end{abstract}


%
\IEEEpeerreviewmaketitle

\newtheorem{Theorem}{Theorem}
\newtheorem{Lemma}{Lemma}
\newtheorem{Proposition}{Proposition}
\newtheorem{Definition}{Definition}
\newtheorem{Example}{Example}

\section{Introduction}

As  a vital component of the Internet of Things (IoS), future backscatter radio frequency identification (RFID) is required to be more reliable and data intensive, and have longer operating range. Although backscatter RFID enjoys low hard complexity and  longer life expectancy, its physical channel experiences deeper fading than conventional one-way channels and leads to severe performance degradation. To mitigate such drawback, many efforts have been made \cite{Ingram2001, Griffin2008, Griffin2009, Langwieser2010, DoYunKim2010,Bauernfeind2012, Denicke2012, Trotter2012, He 2011, He2012, Arnitz2013J, Zheng2012, Hasan2012, AkbarConf2012, He2013, Boyer2013, Karthaus2003, Nikitin2005, Fuschini2008, Xi2009, Bletsas2010, Chakraborty2011,Thomas2012Conf, Thomas2012, Kimionis2012, Boyer2012, Arnitz2013, Griffin2009B}, among which employing multiple antennas for both tags and readers appears to be one of the practical solutions. Such multiple-input multiple-output (MIMO) systems had a great success in mobile communications \cite{Tarokh1998, Tarokh1999, Sandhu2000, Zheng2003, Tse2005} and were also investigated and found promising in backscatter RFID \cite{Ingram2001, Griffin2008, Griffin2009, Langwieser2010, DoYunKim2010,Zheng2012, Hasan2012, AkbarConf2012, Denicke2012, Trotter2012}.


A general $M \times L \times N$ MIMO backscatter RFID channel consists of $M$ reader query antennas, $L$ tag antennas, and $N$ receiving antennas, which can be modeled as a two-way channel with forward
sub-channels and backscattering sub-channels \cite{Ingram2001} \cite{Kim2003} \cite{Griffin2008}, as shown in Fig. \ref{Fig: MIMORFIDBigPic}. In both analytical studies \cite{Ingram2001} \cite{Griffin2008} and real experiments \cite{Griffin2009}, MIMO settings  were shown to be able to  mitigate the channel fading for backscatter RFID. In addition, MIMO settings were shown to be able to increase the  incident power to the tag from the reader transmitting antennas \cite{Hasan2012}, and also increase the operational range \cite{AkbarConf2012} \cite{DoYunKim2010} for backscatter RFID.
Other interesting research including \cite{Denicke2012}, where a method for the determination of the channel coefficients between all antennas was presented; \cite{Trotter2012}, where a hardware design of the multi-antenna tag at $5.8$ GHz are showcased; and \cite{Langwieser2010}, where the researchers described a developed analog frontend for an RFID rapid prototyping system which allows various real-time experiments to investigate MIMO techniques.

The performance of the channel has been also investigated analytically.
Under the quasi-static and the Rayleigh fading assumptions for both the sub-channels, it was shown that for the $M \times L \times N$ backscatter channel, the diversity order achieves $\min(N,L)$ for the uncoded case \cite{He2012}, and the diversity order achieves $L$ for the orthogonal space-time coded case \cite{He2013} \cite{Boyer2013}. Moreover, the diversity order cannot be greater than $L$ \cite{Boyer2013}.
Given the results from the above literatures, diversity order of $L$ seems to be the fundamental limit of the $M \times L \times N$ backscatter channel. However, this is only the case when the conventional uniform query is employed at the query end. Very recently,  \cite{He2014UnitaryQuery} showed that uniform query actually cannot take the advantages of the multiple reader transmitting antennas, and for the first time, \cite{He2014UnitaryQuery} introduced \emph{query diversity } by proposing the \emph{unitary query}, which can considerably improve the performance of the $M \times L \times N$ backscatter channel and can make the diversity order of the backscatter channel much larger than $L$.
Exploiting query diversity is an emerging research direction for high performance backscatter RFID.

\subsection{Motivations: Incorporating the classical OSTBC with the recently proposed unitary query}
Before unitary query was proposed in \cite{He2014UnitaryQuery}, employing orthogonal space-time block code (OSTBC) at the tag end had been investigated for backscatter RFID \cite{Boyer2013, He2013, Zheng2012}. Due to its low complexity (in the sense of encoding and decoding) and high performance, this classical OSTBC is widely adopted in industrial standards and also very attractive for backscatter RFID  which is generally power and hardware limited.  Incorporating classical OSTBC with the recently proposed unitary query seems to be a promising solution for future backscatter RFID to achieve high performance while keeping low complexity. However, in this paper we will first show that there is a linear decoding problem for simple, direct employment of OSTBC together with unitary query and that this decoding problem will lead to performance degradation.


 Hence novel ideas are needed to jointly take advantages of OSTBC and query diversity. We thus propose block-level unitary query and the corresponding modified OSTBC, and refer this novel design strategy the \emph{BUTQ-mOSTBC design pair}. It is worth emphasizing that this design pair is particularly proposed for the $M \times L \times N$ MIMO backscatter RFID channel by a unique marriage between the very recently proposed unitary query and the classical OSTBC. We will show that the proposed BUTQ-mOSTBC can address the linear decoding problem of directly employing OSTBC with unitary query, and can keep the simplicity (easy to encode and decode) and high performance properties (full diversity) of the classical OSTBC.

\subsection{Contributions}
The major contributions of this work include:
\begin{itemize}
\item For the $M \times L \times N$ MIMO backscatter RFID channel of particular interest here, we propose the \emph{block-level unitary query} by extending the unitary query scheme and propose the corresponding modified orthogonal space-time block code. Such a novel \emph{BUTQ-mOSTBC} allows linear decoding and can achieve the potential of the query diversity of the $M \times L \times N$ backscatter channel for OSTBC.
  \item We derive a linear decoder for the proposed BUTQ-mOSTBC design pair and derive the closed-form expression of the asymptotic symbol error rate (SER) expression and diversity order for the design pair.
\end{itemize}

This paper is organized as follows: In Section \ref{Sec: Channel_Model}, we give a brief introduction of the $M \times L \times N$ MIMO backscatter RFID channel, describe the unitary query idea recently proposed for achieving query diversity (or time diversity) for this specific channel, and show that there is a decoding problem for the unitary query when OSTBC is employed directly. In Section \ref{Sec: BUTQ-mOSTBC}, we propose the BUTQ-mOSTBC design pair, present a linear decoder for the proposed BUTQ-mOSTBC, and derive the closed-form SER performance expression and the diversity order. In Section \ref{Sec: Simulations}, we conduct Monte Carlo simulations and discuss the simulation results. Finally we summarize this work in Section \ref{Sec: Conclusion}.


\emph{Notations}: In this paper,  $\mathbb{Q}(\cdot)$ means the $Q$ function; $\mathbb{E}_X(\cdot)$,  $\|\cdot\|$, $(\cdot)^T$, and $(\cdot)^*$ the expectation over the density of $X$,  the magnitude of a complex number,  the transpose of a matrix, and the conjugate of a complex number, respectively;  $a \propto b$ means that $a$ is proportional to $b$, and $\min(c,d)$ means the minimum of $c$ and $d$.


\begin{table*}[!t]
\centering
\caption{Design Pair Abbreviations.}
\begin{tabular}{|c|c|c|}
\hline
Design Pair        & Query End        & Tag End               \\
\hline\hline
UFQ-STC            & uniform query    &  space-time coding    \\
\hline
UFQ-OSTBC          & uniform query    &  orthogonal space-time block coding                   \\
\hline
UFQ-Alamouti       & uniform query    &  Alamouti's code                  \\
\hline
UTQ-STC            & unitary query    &  space-time coding                            \\
\hline
UTQ-OSTBC          & unitary query    &  orthogonal space-time block coding                \\
\hline
UTQ-Alamouti       & unitary query    &  Alamouti's code                  \\
\hline
BUTQ-mOSTBC        & block-level unitary query          & modified orthogonal space-time block coding                 \\
\hline
BUTQ-mAlamouti       & block-level unitary query      &  modified Alamouti's code                  \\
\hline
\end{tabular}\label{Tab: CorrelationAffect Comparision}
\end{table*}

\section{The $M \times L \times N$ Channel and Unitary Query}\label{Sec: Channel_Model}
\subsection{The $M \times L \times N$ MIMO Backscatter RFID Channel}
\begin{figure}
\centering
  \includegraphics[scale=0.6]{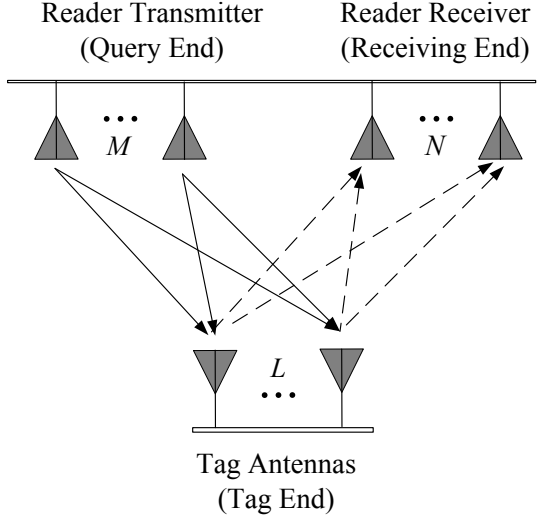}\\
  \caption{The $M \times L \times N$ backscatter channel. The channel consists three operational ends: the query end (with $M$ query antennas), the tag end (with $L$ tag antennas) and the receiving end (with $N$ receiving antennas). The query antennas transmit  unmodulated (query) signals to the RF tag and the RF tag scatters a modulated signal back to the reader.}\label{Fig: MIMORFIDBigPic}
\end{figure}

As shown in Fig. \ref{Fig: MIMORFIDBigPic}, the signal-channel structure of the $M \times L \times N$ MIMO backscatter RFID channel \cite{ Griffin2008, Boyer2013, He2014UnitaryQuery} in a quasi-static fading condition can be described by:
\begin{align}\label{Eq: RFID_Channel_Model}
\mathbf{R}=((\mathbf{Q}\mathbf{H})\circ\mathbf{C})\mathbf{G}+\mathbf{W},
\end{align}
where $\mathbf{Q}$ is the query matrix (with size $T \times M$), representing the query signals sent from the the $M$ reader query (transmitting) antennas to the tag over $T$ time slots (i.e. $T$ symbol times); $\mathbf{H}$
is the channel gain matrix (with size $M \times L$) from the reader transmitter to the tag, representing the forward sub-channels; $\mathbf{C}$ is the coding matrix (with size $T \times L$), where the tag transmits coded or uncoded symbols from its $L$ antennas over $T$ time slots; $\mathbf{G}$ is the channel gain matrix (with size $L \times N$) from the tag to the reader receiver, representing the backscattering sub-channels. Finally the received signals at $N$ reader receiving antennas over $T$ time slots are represented by matrix $\mathbf{R}$ with size $T \times N$, and $\mathbf{W}$ is with the same size as that of $\mathbf{R}$, representing the noises at the $N$ reader receiving antennas over $T$ time slots. Here $\circ$ is the Hadamard product. Typically, both $\mathbf{H}$ and $\mathbf{G}$ are modeled as full rank matrices with i.i.d complex Gaussian entries, and $\mathbf{W}$ is additive white Gaussian noise (AWGN).

The $M \times L \times N$ MIMO backscatter RFID channel, which can be characterized as a \emph{query-fading-coding-fading} structure, is essentially different from conventional one-way MIMO wireless channels: it has one more layer of fading $\mathbf{H}$ and one more signaling mechanism represented by the query matrix $\mathbf{Q}$. In addition, as we can see that the Hadamard product operation in \eqref{Eq: RFID_Channel_Model}, makes the received signals have some non-linear structure, this is due to the backscatter principal.  Because of its special signaling-channel structure, the backscatter RFID channel behaves completely different from that of the one-way channel \cite{Griffin2008, He2013, Boyer2013}. It is also worth mentioning here that, the backscatter channel and the keyhole channel (also has two layers of fading) are also essentially different. The keyhole channel is still a one-way channel, which has only two operational ends (the transmitter and receiver), and the signals will not be reflected back to the receiver, while the backscatter channel has three operational ends and the information to be transmitted is at the middle end (the tag end). In \cite{Boyer2013}, the researchers gave a detail discussion on the essential differences of the two channels. In general, the $M \times L \times N$ MIMO backscatter RFID channel is more complicated than the keyhole channel.

\subsection{A very recent progress: query diversity via unitary query}
There are three operational ends in the $M \times L \times N$ backscatter RFID channel. The diversity schemes at the tag end and the reader receiving end have attracted a lot of attention, and space-time coding and diversity combining techniques have been proposed at the two ends. The query end, however, had been generally ignored for diversity schemes, until unitary query was proposed in \cite{He2014UnitaryQuery} very recently. For unitary query, the query matrix is given by a unitary matrix \cite{He2014UnitaryQuery}:
\begin{align}
\mathbf{Q}\mathbf{Q}^H=\mathbf{I}.
\end{align}
Compared with the conventional uniform query:
\begin{align}
\mathbf{Q}=\frac{1}{\sqrt{M}}\left(
\begin{matrix}
  1 & \cdots & 1 \\
  \vdots & \ddots & \vdots \\
  1 & \cdots & 1
 \end{matrix}\right),
\end{align}
unitary query can create  query diversity (or time diversity)  via multiple query antennas, and hence has great potential to improve the performance of backscatter RFID \cite{He2014UnitaryQuery}.
In this paper, the unitary query employed at the reader query end together with space-time coding employed at the tag end are referred as the UTQ-STC design pair, the abbreviations of other possible joint designs of query signals and coding signals are also listed in Table \ref{Tab: CorrelationAffect Comparision}.

\section{Block Level Unitary Query and Corresponding Modified Orthogonal Space-time Code} \label{Sec: BUTQ-mOSTBC}
Because of the success of the classical OSTBC in conventional wireless channels,
integrating it with the recently proposed unitary query seems to be a promising solution for future backscatter RFID. In this section, we first show that UTQ-OSTBC can only be decoded via exhaustive search and thus cannot fully unitize the diversity potential of the channel, and consequently we propose block-level unitary query at the query end and corresponding modified orthogonal space-time code at the tag end, which is referred as the BUTQ-mOSTBC design pair. BUTQ-mOSTBC can address the linear decoding problem of  UTQ-OSTBC and can fully utilize the query diversity of the channel.

\subsection{Performance of OSTBC with Unitary Query}
We start from considering the case when unitary query and Alamouti's code, the order 2 OSTBC, are employed. This design pair is referred as UTQ-Alamouti, or a order 2 UTQ-OSTBC design. For simplicity we first consider the $2 \times 2 \times 1$ backscatter channel. In this case the received signals are given by
\begin{align}
r_{1,1}& = h_{1,1}g_{1,1}c_1    +h_{1,2}g_{2,1}c_2        +w_{1,1} \nonumber\\
       & = \mathbb{H}_{1}c_1         +\mathbb{H}_{2}c_2             +w_{1,1} \nonumber\\
r_{2,1}& =-h_{2,1}g_{1,1}c_2^*  +h_{2,2}g_{2,1}c_1^*      +w_{2,1} \nonumber\\
       & =-\mathbb{H}_{3}c_2^*       +\mathbb{H}_{4}c_1^*           +w_{2,1},
\end{align}
over two symbol times, where $r_{t,n}$'s, $h_{m,l}$'s, $g_{l,n}$'s and $w_{t,n}$'s are the entries of $\mathbf{R}$, $\mathbf{H}$, $\mathbf{G}$, and $\mathbf{W}$, and we define $\mathbb{H}_{1}\triangleq h_{1,1}g_{1,1}$, $\mathbb{H}_{2}\triangleq h_{1,2}g_{2,1}$, $\mathbb{H}_{3}\triangleq h_{2,1}g_{1,1}$, and $\mathbb{H}_{4}\triangleq h_{2,2}g_{2,1}$.

Clearly, the linear decoder cannot be used to decode the above UTQ-Alamouti design pair, as the OSTBC linear decoder is based on the assumption that the channel does not change in two consecutive symbol times (i.e.,  $\mathbb{H}_{1}=\mathbb{H}_{3}$ and $\mathbb{H}_{2}=\mathbb{H}_{4}$), which are apparently not true for UTQ-Alamouti. Therefore, to decode a UTQ-Alamouti, exhaustive search has to be employed (i.e., comparing each code word in the code book and choose the one having the minimum distance with the received signal). Recall that Alamouti's code is given by the coding matrix
\begin{align}
\mathbf{C}=
\begin{pmatrix}
  c_{1} &    c_{2}  \\
  -c_{2}^* & c_{1}^*  \\
\end{pmatrix}.
\end{align}
For the binary shift-keying (BPSK) case, three possible error determinations are
\begin{align}
\begin{pmatrix}
  c_{1} &    c_{1}  \\
  -c_{1}^* & c_{1}^*  \\
\end{pmatrix},
\begin{pmatrix}
  c_{2} &    c_{2}  \\
  -c_{2}^* & c_{2}^*  \\
\end{pmatrix},
\begin{pmatrix}
  c_{2} &    c_{1}  \\
  -c_{1}^* & c_{2}^*  \\
\end{pmatrix}.
\end{align}
The performance is determined by the code difference matrix between the transmitted code word and the wrong code word having the shortest distance to the transmitted code word, i.e.
\begin{align}
\mathbf{\Delta}=
\begin{pmatrix}
  0 &    c_{2}-c_{1}  \\
  c_{1}^*-c_{2}^* & 0  \\
\end{pmatrix}.
\end{align}
Using the performance measure given in Theorem 1 of \cite{He2014UnitaryQuery}, the performance measure for the UTQ-Alamouti is given by
\begin{align}
R_{unitary}=\sum_{t=1}^T\min(N,\|\Delta\|_0)=1+1=2,
\end{align}
which is the same as the performance measure for the UFQ-Alamouti:
\begin{align}
R_{uniform}=\min(N\times rank(\Delta),L)=L=2.
\end{align}
Based on Theorem 1 of \cite{He2014UnitaryQuery}, it means that the UTQ-Alamouti has a similar performance as that of  UFQ-Alamouti. Similarly, we could easily check that the above observations are also applicable to higher order of UTQ-OSTBC designs for the $M \times L \times N$ backscatter RFID channels. So it is clearly that UTQ-OSTBC is not a good way to incorporate the query diversity and OSTBC.

\subsection{Block-level Unitary Query and Corresponding Modified OSTBC: the BUTQ-mOSTBC Design Pair}
In this section, to address the above concerns in the UTQ-OSTBC design pair, we propose block-level unitary query at the query end and present the corresponding modified OSTBC at the tag end, which together are called BUTQ-mOSTBC design pair
for the $M \times L \times N$ MIMO RFID channel.

\begin{Definition}\label{Definition: BlockLevelUnitaryQuery}
A BUTQ-mOSTBC design pair consists of two parts, a block-level unitary query employed at the query end and a modified OSTBC at the tag end. More specifically, in this paper, the block-level unitary query is defined by the query matrix
\begin{align}\label{Eq: BlockLevelUnitaryQuery}
\mathbf{Q}=\mathbf{Q}_0 \otimes \mathbf{1}_M
\end{align}
where $\mathbf{Q}_0$ is a unitary matrix, $\otimes$ is the Kronecker product, and
$\mathbf{1}_M=\underbrace{(1,1,\cdots,1)^T}_{\text{M terms}}$.
The modified space-time code corresponding to the block-level unitary query is given by the coding matrix
\begin{align}\label{Eq: ModifiedSpaceTimeCode}
\mathbf{C}=\left(
           \begin{array}{c}
              \mathbf{C}_1   \\
             \vdots        \\
              \mathbf{C}_M    \\
           \end{array}
         \right),
\end{align}
where $\mathbf{C}_1=\mathbf{C}_2=\cdots=\mathbf{C}_M=\mathbf{C}_0$, with  $\mathbf{C}_0$ being an original OSTBC.
\end{Definition}

We would like to emphasize that the above definition only represents a specific BUTQ-mOSTBC design strategy. The BUTQ-mOSTBC can have many other forms: for instance, any permutation of the rows of the original \eqref{Eq: BlockLevelUnitaryQuery} and \eqref{Eq: ModifiedSpaceTimeCode} will result in a specific BUTQ-mOSTBC design which can provide the same BER performance as the original design. For instance, a BUTQ-mOSTBC design as
\begin{align}
\mathbf{Q}=\left(
           \begin{array}{c}
               \mathbf{Q}_1 \\
               \vdots  \\
              \mathbf{Q}_M
           \end{array}
         \right), \;\;\;\;\;
\mathbf{C}=\mathbf{C}_0 \otimes \mathbf{1}_M
         , \;\;\;\;\;
\end{align}
where $\mathbf{Q}_1=\mathbf{Q}_2=\cdots=\mathbf{Q}_M=\mathbf{Q}_0$, is equivalent to the design pair in Definition \ref{Definition: BlockLevelUnitaryQuery} in terms of the bit error rate (BER) performance.

We use the following examples to further illustrate what is a BUTQ-mOSTBC design pair.

\begin{Example}\label{Ex: BUTQ-mAlamouti}
We consider a $2 \times 2 \times 2$ backscatter RFID channel. Let the Alamouti's code
\begin{align}\label{Eq: MatrixQ}
\mathbf{C_0}=\left(
           \begin{array}{cc}
              c_1 & c_2 \\
              -c_2* & c_1* \\
           \end{array},
         \right)
\end{align}
to be the original code, and let $\mathbf{Q}_0=\bigl(\begin{smallmatrix}1&0\\ 0&1\end{smallmatrix} \bigr)$, hence the BUTQ-mOSTBC design pair is with the block-level unitary query
\begin{align}\label{Eq: BlockLevelUnitaryEx1}
\mathbf{Q}=\mathbf{Q}_0 \otimes \mathbf{1}_M
=\left(\begin{array}{cc}
              1 & 0 \\
              1 & 0 \\
              0 & 1 \\
              0 & 1 \\
           \end{array}\right),
\end{align}
and the corresponding modified Alamouti's code
\begin{align}\label{Eq: ModifiedSTCEx1}
\mathbf{C}
=\left(\begin{array}{c}
              \mathbf{C}_1 \\
              \mathbf{C}_2
           \end{array}
         \right)
=\left(
           \begin{array}{cc}
              c_1 & c_2 \\
              -c_2* & c_1* \\
              c_1 & c_2 \\
              -c_2* & c_1* \\
           \end{array}
         \right).
\end{align}
\end{Example}

\begin{figure}
\centering
  \includegraphics[scale=0.6]{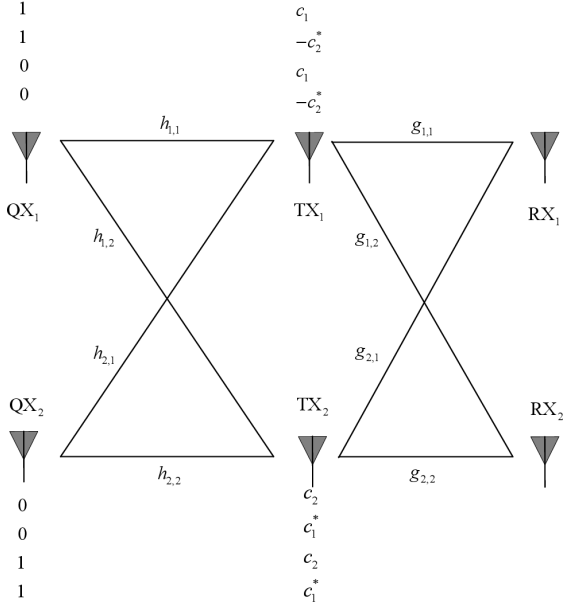}\\
  \caption{The signal flow of the BUTQ-mOSTBC design pair given in Example \ref{Ex: BUTQ-mAlamouti} for the
  $2 \times 2 \times 2$ MIMO backscatter RFID channel.}\label{Fig: SignalFlowBLUQ}
\end{figure}

Table \ref{Tab: BUTQ-mOSTBC_Signals} and Fig. \ref{Fig: SignalFlowBLUQ} show the encoding and the signal flow of the BUTQ-mOSTBC design pair in Example 1 for the $2 \times 2 \times 2$ backscatter RFID channel.

\begin{Example}
Other forms of the block-level unitary query can be obtained via permutations of the rows of \eqref{Eq: BlockLevelUnitaryEx1} and \eqref{Eq: ModifiedSTCEx1} in the same way. For instance, we can have
\begin{align}\label{Eq: BlockLevelUnitaryEx2}
\mathbf{Q}=\left(
           \begin{array}{c}
              \mathbf{Q}_1 \\
              \mathbf{Q}_2
           \end{array}
         \right)
=\left(
           \begin{array}{cc}
              1 & 0 \\
              0 & 1 \\
              1 & 0 \\
              0 & 1 \\
           \end{array}
         \right),
\end{align}
and the corresponding modified space-time code as
\begin{align}\label{Eq: ModifiedSTCEx2}
\mathbf{C}
=\mathbf{C}_0 \otimes \mathbf{1}_M
=\left(
           \begin{array}{cc}
              c_1 & c_2 \\
              c_1 & c_2 \\
              -c_2* & c_1* \\
              -c_2* & c_1* \\
           \end{array}
         \right).
\end{align}
\end{Example}

In the next sub-section, we will show that the proposed BUTQ-mOSTBC design pair can be decoded linearly and can achieve the full diversity potential of the $M \times L \times N$ MIMO backscatter RFID channel.


\begin{table*}[!t]
\centering
\caption{The encoding of the proposed BUTQ-mOSTBC in Example \ref{Ex: BUTQ-mAlamouti} for the $2 \times 2 \times 2$ backscatter RFID channel.}
\begin{tabular}{|c|c|c|c|c|c}
\hline
              & query antenna  1   & query antenna 2 & tag antenna 1       &   tag antenna 2          \\
\hline\hline
$t=1$         &       1            &  0              &      $c_{1}$        &   $c_2$           \\
\hline
$t=2$         &       1            &  0              &      $-c_{2}^*$     &   $c_1^*$  \\
\hline
$t=3$         &       0            &  1              &      $c_{1}$        &   $c_2$           \\
\hline
$t=4$         &       0            &  1              &      $-c_{2}^*$     &   $c_1^*$  \\
\hline
\end{tabular}\label{Tab: BUTQ-mOSTBC_Signals}
\end{table*}

\subsection{Decoding of the BUTQ-mOSTBC}\label{Sec: Decoding}
We now investigate the decoding of the specific BUTQ-mOSTBC in Example \ref{Ex: BUTQ-mAlamouti}, which is referred as the BTUQ-mAlamouti. The signal flow of the transmitting and receiving structure for the $2 \times 2 \times 2$ channel is illustrated in Fig. \ref{Fig: SignalFlowBLUQ}, where the received signals are
\begin{align}
&r_{1,1}= h_{1,1}g_{1,1}c_1+h_{1,2}g_{2,1}c_2+w_{1,1}        \nonumber \\
&r_{2,1}=-h_{1,2}g_{2,1}c_1^*   +h_{1,1}g_{1,1}c_2^*+w_{2,1} \nonumber \\
&r_{3,1}= h_{2,1}g_{1,1}c_1+h_{2,2}g_{2,1}c_2+w_{3,1}        \nonumber \\
&r_{4,1}=-h_{2,2}g_{2,1}c_1^*   +h_{2,1}g_{1,1}c_2^*+w_{4,1}
\end{align}
for the first receiving antenna, and
\begin{align}
&r_{1,2}= h_{1,1}g_{1,2}c_1+h_{1,2}g_{2,2}c_2+w_{1,2} \nonumber\\
&r_{2,2}=-h_{1,2}g_{2,2}c_1^*   +h_{1,1}g_{1,2}c_2^*+w_{2,2}\nonumber\\
&r_{3,2}= h_{2,1}g_{1,2}c_1+h_{2,2}g_{2,2}c_2+w_{3,1}\nonumber\\
&r_{4,2}=-h_{2,2}g_{2,2}c_1^*   +h_{2,1}g_{1,2}c_2^*+w_{4,2}
\end{align}
for the second receiving antenna.

For the $2 \times 2 \times 2$ channel, we process the received signals as following to obtain the combined signals for $c_1$ and $c_2$:
\begin{align}
\tilde{c}_1
=&\underbrace{\overbrace{h_{1,1}^*g_{1,1}^*r_{1,1}-h_{1,2}g_{2,1}r_{2,1}^*}^{\text{first block}}
            + \overbrace{h_{2,1}^*g_{1,1}^*r_{3,1}-h_{2,2}g_{2,1}r_{4,1}^*}^{\text{second block}}}_{n=1}\nonumber\\
+&\underbrace{\overbrace{h_{1,1}^*g_{1,2}^*r_{1,2}-h_{1,2}g_{2,2}r_{2,2}^*}^{\text{first block}}
            + \overbrace{h_{2,1}^*g_{1,2}^*r_{3,2}-h_{2,2}g_{2,2}r_{4,2}^*}^{\text{second block}}}_{n=2}.
\end{align}
After some algebra operations, we have
\begin{align}
\tilde{c}_1
=&\underbrace{(\|h_{1,1}g_{1,1}\|^2+\|h_{1,2}g_{2,1}\|^2+\|h_{2,1}g_{1,1}\|^2+\|h_{2,2}g_{2,1}\|^2)c_1}_{\text{signals, $n=1$}} \nonumber\\
+&\underbrace{h_{1,1}^*g_{1,1}^*w_{1,1}-h_{1,2}g_{2,1}w_{2,1}^*+h_{2,1}^*g_{1,1}^*w_{3,1}-h_{2,2}g_{2,1}w_{4,1}^*}_{\text{noises, $n=1$}} \nonumber\\
+&\underbrace{(\|h_{1,1}g_{1,2}\|^2+\|h_{1,2}g_{2,2}\|^2+\|h_{2,1}g_{1,2}\|^2+\|h_{2,2}g_{2,2}\|^2)c_1}_{\text{signals, $n=2$}} \nonumber\\
+&\underbrace{h_{1,1}^*g_{1,2}^*w_{1,2}-h_{1,2}g_{2,2}w_{2,2}^*+h_{2,1}^*g_{1,2}^*w_{3,2}-h_{2,2}g_{2,2}w_{4,2}^*}_{\text{noises, $n=2$}}. \end{align}
Note that $w_{t,n}$'s are i.i.d. complex Gaussian r.v.s, therefore it is followed that
\begin{align}
&\left(\textmd{noises, } n=1\right)+\left(\textmd{noises, } n=2\right) \nonumber\\
& \overset{\textmd{identically distributed with}} {=}  \sqrt{\sum_{n=1}^2\sum_{t=1}^2\sum_{l=1}^2\|h_{t,l}\|^2 \|g_{l,n}\|^2}w'
\end{align}
where $w'$ is a unity variance complex Gaussian noise.
Therefore
\begin{align}
\tilde{c}_1=&\underbrace{\left(\sum_{n=1}^2\sum_{t=1}^2\sum_{l=1}^2\|h_{t,l}\|^2\|g_{l,n}\|^2\right)c_1}_{\text{signal term}} \nonumber \\
            &+\underbrace{\sqrt{\sum_{n=1}^2\sum_{t=1}^2\sum_{l=1}^2\|h_{t,l}\|^2 \|g_{l,n}\|^2}w'}_{\text{noise term}}.
\end{align}
Similarly, the combined signal for $c_2$ is given by
\begin{align}
\tilde{c}_2=&\underbrace{\overbrace{h_{1,2}^*g_{2,1}^*r_{1,1}-h_{1,1}g_{1,1}r_{2,1}^*}^{\text{first block}}+\overbrace{h_{2,2}^*g_{2,1}^*r_{3,1}-h_{2,1}g_{1,1}r_{4,1}^*}^{\text{second block}}}_{n=1}\nonumber\\
           +&\underbrace{\overbrace{h_{1,2}^*g_{2,2}^*r_{1,2}-h_{1,1}g_{1,2}r_{2,2}^*}^{\text{first block}}+\overbrace{h_{2,2}^*g_{2,2}^*r_{3,2}-h_{2,1}g_{1,2}r_{4,2}^*}^{\text{second block}}}_{n=2},
\end{align}
and after some algebra we have
\begin{align}
\tilde{c}_2=&\underbrace{\sum_{n=1}^2\sum_{t=1}^2\sum_{l=1}^2\|h_{t,l}\|^2 \|g_{l,n}\|^2c_2\bar{\gamma}}_{\textmd{signal term}} \nonumber\\
            &+\underbrace{\sqrt{\sum_{n=1}^2\sum_{t=1}^2\sum_{l=1}^2\|h_{t,l}\|^2 \|g_{l,n}\|^2}w''}_{\textmd{noise term}},
\end{align}
where $w''$ is another unity variance complex Gaussian noise.
Now we define the following metric between $\tilde{c}_{1}$ and a symbol $s_{i}$,
\begin{align}
d(\tilde{c}_{1},s_{i})=\left|\tilde{c}_{1}-s_{i}\sum_{n=1}^2\sum_{t=1}^2\sum_{l=1}^2\|h_{t,l}\|^2 \|g_{l,n}\|^2\right|,
\end{align}
the decoder choose $s_i$ for $c_1$ iff
\begin{align}
d(\tilde{c}_{1},s_{i})\le d(\tilde{c}_{1},s_{k}), \;\;\;\;\;\;\;\;\; \forall i\ne k.
\end{align}
Similarly, the decoder choose $s_i$ for $c_2$ iff
\begin{align}
d(\tilde{c}_{2},s_{i})\le d(\tilde{c}_{2},s_{k}), \;\;\;\;\;\;\;\;\; \forall i\ne k.
\end{align}

Using $Z$ to denote the instantaneous SNR, we can show that, for $c_1$ and $c_2$, the instantaneous SNR is given by
\begin{align}
Z
=\left(\frac{\textmd{signal term}}{\textmd{noise term}}\right)^2
=\bar{\gamma}\sum_{n=1}^2\sum_{t=1}^2\sum_{l=1}^2\|h_{t,l}\|^2 \|g_{l,n}\|^2,
\end{align}
where $\bar{\gamma}$ is the average SNR.

This decoding process can be easily generalized to a more general, higher order BUTQ-mOSTBC. It can also be checked that, for the general $M \times L \times N$ backscatter RFID channel, the instantaneous SNR using a similar decoding process can be expressed as
\begin{align}
Z
=\bar{\gamma}\sum_{n=1}^N\sum_{t=1}^M\sum_{l=1}^L\|h_{t,l}\|^2 \|g_{l,n}\|^2.
\end{align}
Intuitively the above instantaneous SNR is very diversified since it includes all channel paths in the $M \times L \times N$ channel.

\subsection{Performance Analysis}\label{Sec: BUTQ-mOSTBC Performance Analysis}
We now proceed to analytically study the performance of the proposed BUTQ-mOSTBC for the $M \times L \times N$ channel.
The instantaneous SNR can be written as
\begin{align}
Z=\bar{\gamma}\sum_{n=1}^N\sum_{t=1}^M\sum_{l=1}^L\|h_{t,l}\|^2 \|g_{l,n}\|^2
=\sum_{l=1}^LZ_{l}
\end{align}
where $Z_l$ is defined as
\begin{align}
Z_l\triangleq \bar{\gamma}\sum_{n=1}^N\sum_{t=1}^M\|h_{t,l}\|^2 \|g_{l,n}\|^2,
\end{align}
and it can be shown that $Z_l$'s are independent given that $h_{t,l}$'s and $g_{l,n}$'s are independent.
The asymptotic symbol error rate (SER) in a closed-form is therefore given by
\begin{align}\label{Eq: SER_Asym}
\mathbb{P}(\bar{\gamma})&=\nonumber \\
  &\left\{
  \begin{array}{l l l}
  \frac{\Gamma(1/2+LN)}{2\sqrt{\pi}\Gamma(1+LN)}\left(\frac{\Gamma(M-N)}{\Gamma(M)}\right)^L(g\bar{\gamma})^{-LN}, & \quad \text{if $N<M$};\\
    &\\
  \frac{\Gamma(1/2+LN)}{2\sqrt{\pi}\Gamma(1+LN)}\left(\frac{\ln(g\bar{\gamma})}{\Gamma(N)}\right)^L(g\bar{\gamma})^{-LN}, & \quad \text{if $N=M$};\\
    &\\
  \frac{\Gamma(1/2+LM)}{2\sqrt{\pi}\Gamma(1+LM)}\left(\frac{\Gamma(N-M)}{\Gamma(N)}\right)^L(g\bar{\gamma})^{-LM}, &  \quad \text{if $N\geq M$},
  \end{array} \right.
\end{align}
where $g$ is a constant depending the modulation being used. The detail derivation is given in the Appendix.


From \eqref{Eq: SER_Asym}, we can see that the diversity order for the BUTQ-mOSTBC can achieve
\begin{align}\label{Eq: BUTQ-mOSTBC_DiversityOrder}
d_{BUTQ-mOSTBC}=L\min(M,N).
\end{align}
Recall that the diversity order for UFQ-OSTBC with linear decoder is given by \cite{Boyer2013} \cite{He2012}:
\begin{align}\label{Eq: UFQ-OSTBC_DiversityOrder}
d_{UFQ-OSTBC}=L,
\end{align}
and it can be shown that diversity order of the UTQ-OSTBC with exhaustive search is the same as that of UFQ-OSTBC, i.e.,
\begin{align}\label{Eq: UFQ-OSTBC_DiversityOrder}
d_{UTQ-OSTBC}=L.
\end{align}
It suggests that the BUTQ-mOSTBC can yield much better performance than that of the UTQ-OSTBC and the UFQ-OSTBC asymptotically. Table \ref{Tab: DiversityOrderandDecoding} compares the achievable diversity orders and decoding approaches for different design pairs.

We can see that the data rate of the BUTQ-mOSTBC design pair is $\frac{1}{M}$ of that of the UFQ-OSTBC if the same modulation is used, and thus leads to a data rate loss. Fortunately, this data rate loss can be compensated by the diversity gain of the proposed BUTQ-mOSTBC.
To have the same data rate (bit rate), a higher order of modulation can be used, which will result in a smaller constellation size, and consequently the diversity gain term of the BUTQ-mOSTBC becomes
\begin{align}
(\tilde{g}\bar{\gamma})^{-L\min(M,N)},
\end{align}
where $\tilde{g}<g$ is a constant which depends on the higher modulation being used, and we have
\begin{align}\label{Eq: HigherOrderModulation}
\lim_{\bar{\gamma} \rightarrow \infty} \frac{\mathbb{P}_{\text{BUTQ-mOSTBC}}(\bar{\gamma})}{\mathbb{P}_{\text{UFQ-OSTBC}}(\bar{\gamma})}
\propto \lim_{\bar{\gamma} \rightarrow \infty}
\frac{(\tilde{g}\bar{\gamma})^{-L\min(M,N)}}{(g\bar{\gamma})^{-L}} \rightarrow 0.
\end{align}
Therefore \eqref{Eq: HigherOrderModulation} shows that, even with the same bit rate,  the BUTQ-mOSTBC always outperforms the UFQ-OSTBC and the UTQ-OSTBC asymptotically.

\begin{table*}[!t]
\centering
\caption{Achievable Diversity Order and Decoding of Design Pairs.}
\begin{tabular}{|c|p{5.5cm}|p{5cm}|}
\hline
Design Pair               & Diversity Order       & Decoding            \\
\hline\hline
UFQ-STC        & $\leq L$, from \cite{Boyer2013}  &  cannot be linearly decoded in general    \\
\hline
UFQ-OSTBC       & $L$,  from \cite{Boyer2013} \cite{He2013}  &  can be linearly decoded                 \\
\hline
UFQ-Alamouti       & $2$, from \cite{Boyer2013} \cite{He2013}  &  can be linearly decoded                  \\
\hline
UTQ-STC        & not known in general, can be much larger than $L$ with proper design, from \cite{He2014UnitaryQuery}   &  cannot be linearly decoded in general                            \\
\hline
UTQ-OSTBC       & $L$, from this paper   &  cannot be linearly decoded                \\
\hline
UTQ-Alamouti       & $2$, from  this paper  &  cannot be linearly decoded                 \\
\hline
BUTQ-mOSTBC     &  $L \times \min(M,N)$, from this paper         & can be linearly decoded                  \\
\hline
BUTQ-mAlamouti       & $2 \times \min(M,N)$, from this paper      &  can be linearly decoded                 \\
\hline
\end{tabular}\label{Tab: DiversityOrderandDecoding}
\end{table*}

\section{Numerical Simulations and Discussions}\label{Sec: Simulations}
In this section, we perform Monte Carlo simulations and compare the results of the proposed BUTQ-mOSTBC and other two design pairs. We can see that the BUTQ-mOSTBC is a promising design for backscatter RFID systems, due to its high performance and relatively simple coding and decoding methods. In the simulations, we use the same channel model as in previous real measurements \cite{Kim2003} \cite{Griffin2009} and analytical studies \cite{Griffin2008, He2012, He2013, Boyer2013}: the entries of both $\mathbf{H}$  and those of $\mathbf{G}$  follow i.i.d complex Gaussian distribution with zero mean and unity variance. In addition, $\mathbf{H}$ and $\mathbf{G}$ are independent, and the fading is quasi-static. The number of channel realizations in the simulations is adaptive to the SER level, i.e., the simulation will stop if $50$ errors occur. For instance, if the SER is $10^{-3}$, about $\frac{50}{10^{-3}}=5\times 10^{4}$ channel realizations are generated.

\subsection{Asymptotic Closed-Form SERs}
\begin{figure}
\centering
  \includegraphics[scale=0.66]{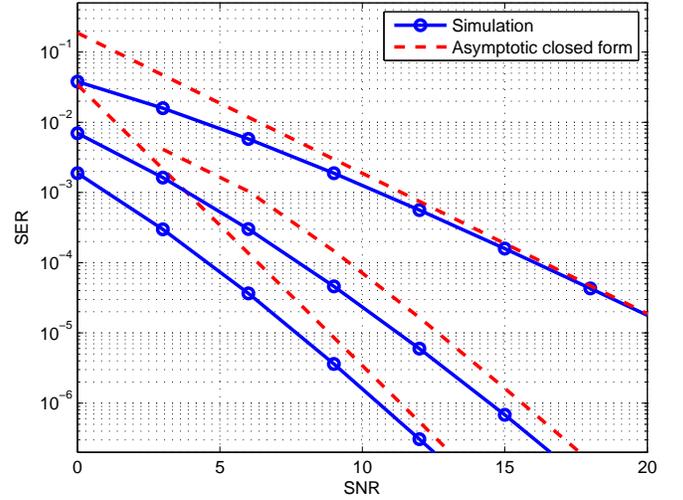}\\
  \caption{Comparisons between the analytical results (asymptotic) and the Monte Carlo simulations for the proposed BUTQ-mOSTBC in MIMO backscatter RFID channels, where BPSK modulation is used. From the top to the bottom: $2 \times 2 \times 1$ channel, $2 \times 2 \times 2$ channel, $2 \times 2 \times 3$ channel.}\label{Fig: CloseFormBUTQ_all}
\end{figure}

We first verify the derived asymptotic closed-formÿSER results by simulations.
As we can see from Fig. \ref{Fig: CloseFormBUTQ_all}, the analytical SERs match well with the simulation results asymptotically. Based on Table \ref{Tab: DiversityOrderandDecoding} , the diversity orders of the proposed BUTQ-mOSTBC for the $2 \times 2 \times 1$, $2 \times 2 \times 2$, and $2 \times 2 \times 3$ channels, are $2$, $4$, and $4$, respectively, which are confirmed by the SER curves in Fig. \ref{Fig: CloseFormBUTQ_all}: the curves of the $2 \times 2 \times 2$ channel and the $2 \times 2 \times 3$ channel are parallel and they are steeper than the curve of the $2 \times 2 \times 1$ channel. Although both the $2 \times 2 \times 2$ and $2 \times 2 \times 3$ channels achieve the same diversity order, the $2 \times 2 \times 3$ channel can provide a considerable better performance. This is due to the $\ln(\bar{\gamma})$ term of the SER expression in \eqref{Eq: SER_Asym} when $M=N$.

\subsection{Performance Comparisons}
We now compare the performance of the proposed BUTQ-mOSTBC design pair with those of the UFQ-OSTBC and the UTQ-OSTBC.
Both the BUTQ-mOSTBC and the UFQ-OSTBC can be decoded linearly, while the UFQ-OSTBC can only be decoded via exhaustive search, despite that the tag end employs OSTBC. For readers' reference, the linear decoder for the BUTQ-mOSTBC is derived in Section \ref{Sec: Decoding}.

To make fair comparisons, the BUTQ-mOSTBC should transmit the same data rate as that of the UFQ-OSTBC and the UTQ-OSTBC, hence a higher modulation should be used in the BUTQ-mOSTBC. In our simulations, for $M=2$, UFQ-OSTBC and UFQ-OSTBC employ BPSK, while the BUTQ-mOSTBC employs quadrature phase-shift keying (QPSK). Thus the data rates are all $1$ bit per symbol time in three schemes.  In addition, for QPSK, the symbol error rate is converted to the bit error rate.
From the simulations, we can see that, even with the same data rate (or equivalently bit rate), significant gains can be brought by the BUTQ-mOSTBC, which is consistent with the analysis given in Section \ref{Sec: BUTQ-mOSTBC Performance Analysis}. We can also observe that, as expected, the UTQ-OSTBC and UFQ-OSTBC achieve the same diversity order and have similarly performances, as we expected. This observation suggests that, although the signal-channel structure of the UTQ-OSTBC is more diverse than that of the UFQ-OSTBC, due to the decoding problem, the potential of the query diversity cannot be fully utilized in the UTQ-OSTBC. Fortunately, the proposed BUTQ-mOSTBC can resolve the decoding problem in the UTQ-OSTBC and thus fully utilize the query diversity advantage.

\begin{figure}
\centering
  \includegraphics[scale=0.66]{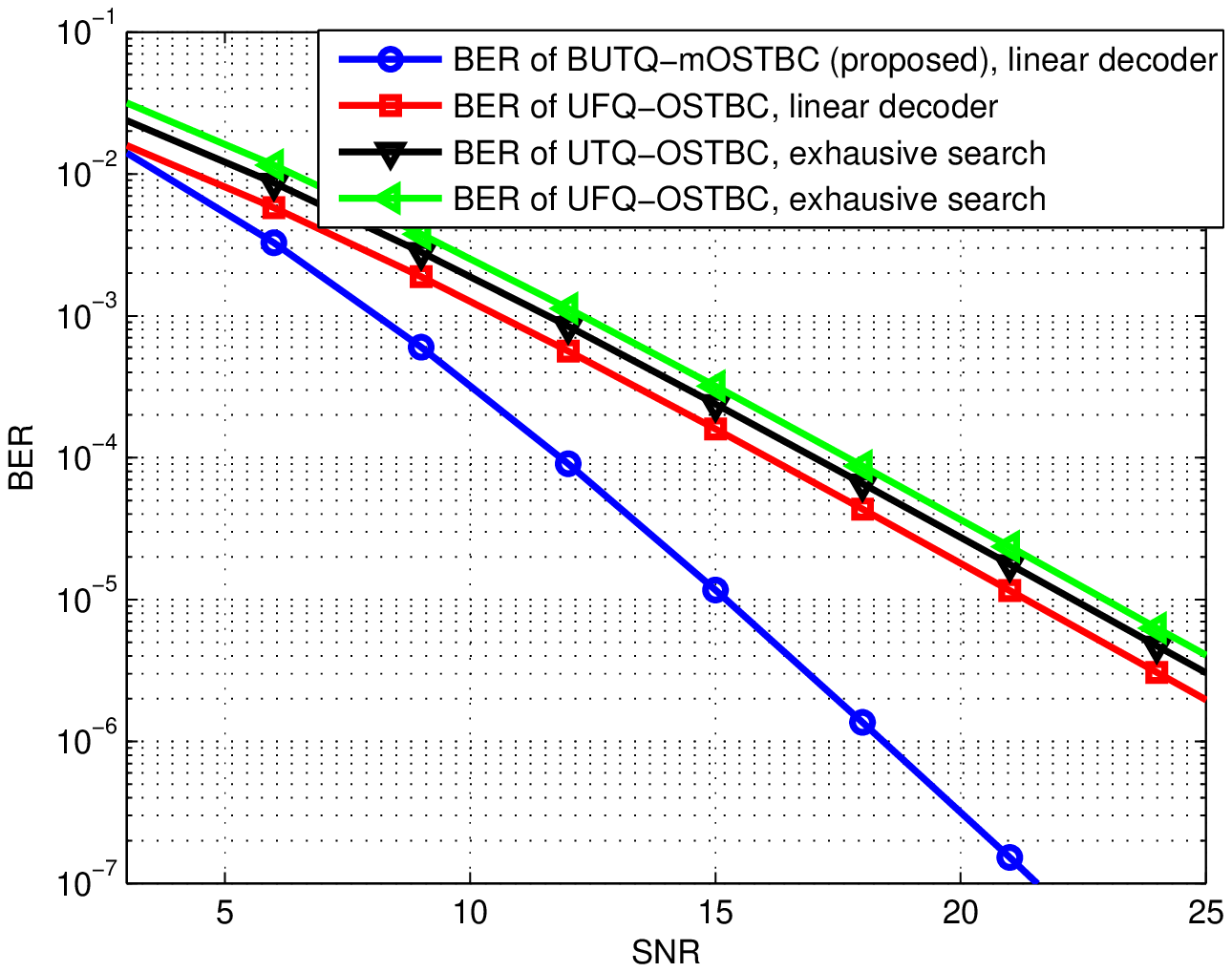}\\
  \caption{Performance comparisons between the proposed BUTQ-mOSTBC, the UTQ-OSTBC and the UFQ-OSTBC for the $2 \times 2 \times 2$ backscatter RFID channel. The data rates for all three methods are 1 bit per symbol time: BUTQ-mOSTBC employs QPSK, and UTQ-OSTBC and UFQ-OSTBC employ BPSK.}\label{Fig: BER_BLUQ_M2L2N2SameBitRate}
\end{figure}

\begin{figure}
\centering
  \includegraphics[scale=0.66]{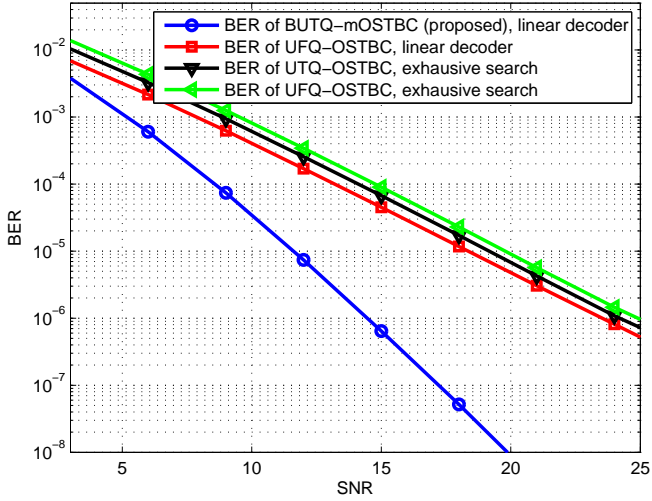}\\
  \caption{Performance comparisons between the proposed BUTQ-mOSTBC, the UTQ-OSTBC and the UFQ-OSTBC for the $2 \times 2 \times 3$ backscatter RFID channel. The data rates for all three design pairs are 1 bit per symbol time: BUTQ-mOSTBC employs QPSK, and UTQ-OSTBC and UFQ-OSTBC employ BPSK.}\label{Fig: BER_BLUQ_M2L2N3SameBitRate}
\end{figure}

\section{Conclusion}\label{Sec: Conclusion}
In this paper, we have proposed  block-level unitary query and its corresponding modified OSTBC, referred as the BUTQ-mOSTBC design pair, for the $M \times L \times N$ MIMO backscatter RFID channel. BUTQ-mOSTBC is a re-design and improvement of the unitary query scheme, which was proposed very recently in \cite{He2014UnitaryQuery}, to tackle the linear decoding problem and the potential performance degradation from directly employing OSTBC together with the unitary query.
BUTQ-mOSTBC can be decoded linearly by using our proposed decoding method.
The closed-form expression of the asymptotic SER shows that the proposed BUTQ-mOSTBC has a diversity order of
$L \times \min(M,N)$, which is larger than the diversity orders of the UFQ-OSTBC and the UTQ-OSTBC. Simulation results confirm the analytical results and show that, when transmitting at the same data rate, the proposed BUTQ-mOSTBC design pair outperforms both the UFQ-OSTBC and the UTQ-OSTBC with significant gains.


%

\section{Appendix}
The SER for the $M \times L \times N$ channel can be obtained by
\begin{align}
\mathbb{P}(\bar{\gamma})=\mathbb{E}_{Z}\left(\mathbb{Q}\left(\sqrt{2Z}\right)\right),
\end{align}
Using the alternative representation of the $Q$ function, and since $Z_l$'s are independent we have
\begin{align}
\mathbb{P}(\bar{\gamma})
&=\mathbb{E}_{Z}\left(\frac{1}{\pi}\int_{\theta=0}^{\pi/2}\exp\left(-\frac{Z}{\sin^2\theta}\right)d\theta\right) \nonumber\\
&=\mathbb{E}_{Z}\left(\frac{1}{\pi}\int_{\theta=0}^{\pi/2}\exp\left(-\frac{\sum_{l=1}^LZ_{l}}{\sin^2\theta}\right)d\theta\right)\nonumber\\
&=\frac{1}{\pi}\int_{\theta=0}^{\pi/2}\prod_{l}^L\mathbb{E}_{Z_l}\left(\exp\left(-\frac{Z_{l}}{\sin^2\theta}\right)\right)d\theta.
\end{align}

\begin{align}\label{Eq: ExpEZl}
&\mathbb{E}_{Z_l}\left(\exp\left(-\frac{Z_l}{\sin^2\theta}\right)\right) \nonumber\\
&=\mathbb{E}\left(\exp\left(-\bar{\gamma}\frac{\sum_{n=1}^N\sum_{t=1}^T\|h_{t,l}\|^2 \|g_{l,n}\|^2}{\sin^2\theta}\right)\right)\nonumber\\
\end{align}
the above expectation in \eqref{Eq: ExpEZl} has been well studied in \cite{He2012} and has the following asymptotic expression

\begin{align}\label{Eq: MIMO_BPSK_Asym}
  &\mathbb{E}\left(\exp\left(-\bar{\gamma}\frac{\sum_{n=1}^N\sum_{t=1}^M\|h_{t,l}\|^2 \|g_{l,n}\|^2}{\sin^2\theta}\right)\right) \nonumber\\
  &\doteq \left\{
  \begin{array}{l l l}
    \left(\sin^{2N}\theta\right)\left(\frac{\Gamma(M-N)}{\Gamma(M)}\right)(g\bar{\gamma})^{-N},
    & \quad \text{if $N<M$};\\
    \left(\sin^{2N}\theta\right)\left(\frac{\ln(g\bar{\gamma})}{\Gamma(N)}\right)(g\bar{\gamma})^{-N}
    & \quad \text{if $N=M$};\\
    \left(\sin^{2M}\theta\right)\left(\frac{\Gamma(N-M)}{\Gamma(N)}\right)(g\bar{\gamma})^{-M},
    &  \quad \text{if $N\geq M$},\\
  \end{array} \right.
\end{align}
therefore
\begin{align}\label{Eq: MIMO_BPSK_Asym}
  &\prod_{l}^L\mathbb{E}_{Z_l}\left(\exp\left(-\frac{Z_{l}}{\sin^2\theta}\right)\right) \nonumber\\
  &\doteq \left\{
  \begin{array}{l l l}
    \left(\sin^{2LN}\theta\right)\left(\frac{\Gamma(M-N)}{\Gamma(M)}\right)^L(g\bar{\gamma})^{-LN},
    & \quad \text{if $N<M$};\\
    \left(\sin^{2LN}\theta\right)\left(\frac{\ln(g\bar{\gamma})}{\Gamma(N)}\right)^L(g\bar{\gamma})^{-LN}
    & \quad \text{if $N=M$};\\
    \left(\sin^{2LM}\theta\right)\left(\frac{\Gamma(N-M)}{\Gamma(N)}\right)^L(g\bar{\gamma})^{-LM},
    &  \quad \text{if $N\geq M$},\\
  \end{array} \right.
\end{align}
and by integrating over $\theta$
\begin{equation}\label{Eq: MIMO_BPSK_Asym}
  \mathbb{P}(\bar{\gamma})
  =\frac{1}{\pi}\int_{\theta=0}^{\pi/2}\prod_{l}^L\mathbb{E}_{Z_l}\left(\exp\left(-\frac{Z_{l}}{\sin^2\theta}\right)\right)d\theta,
\end{equation}
the asymptotic performance given in \eqref{Eq: SER_Asym} can be obtained accordingly.

\ifCLASSOPTIONcaptionsoff
  \newpage
\fi



%

\bibliographystyle{IEEEtran}
\bibliography{RFIDarticle2}

\end{document}